\documentclass[12pt]{revtex4}
\usepackage{epsfig}
\usepackage{graphicx}
\begin{document}
\title{Relativistic Scattering with a Singular Potential in the Dirac Equation}

\author{M Loewe\footnote{Email Address: mloewe@fis.puc.cl} and S Mendizabal\footnote{Email address: smendizabal@fis.puc.cl}}
\address{
Facultad de F\'{\i}sica, Pontificia Universidad Cat\'olica
de Chile, Casilla 306, Santiago 22, Chile}

\begin{abstract}
An elementary treatment of the
Dirac equation in the presence of a three dimensional spherically
symmetric delta potential is presented. We show how to calculate
the cross section using the relativistic wave expansion method for
a one delta potential and two concentric delta potentials. We
compare our results with the cross section calculated in the Born
approximation.
\end{abstract}

\pacs{03.65.Nk,03.65.Pm}

\maketitle

\section{Introduction}

The non relativistic scattering theory, and in particular the
partial wave expansion method, is a very well known issue. An
extensive literature \cite{cohen} exists in this subject. In the
relativistic case, however, the scattering problem with a
potential, has almost not been discussed. This has to do,
probably, with the fact that relativistic quantum mechanics is, in
fact, a theory of many particles and, therefore, quantum field
theory is a more appropriate language to discuss these kind of
problems.

We think, however, that the scattering problem in the frame of the
Dirac equation deserves some attention. This fact has motivated us
to make use of a simple mathematical method, valid for central
potentials, the so called \emph{relativistic partial wave
expansion method} provides us with valuable information for
relativistic particles near the low energy limit, giving us also
values for the phase shifts and cross section of the scattered
wave for different channels of the angular momentum. This method
will be used to complement the work done in \cite{loewe}, a
discussion for bound states in the presence of a delta potential,
which shows how to fix the boundary conditions for the wave
function when crossing the $V(r)=a\delta(r-r_{0})$ potential. We
will like to remark that there are general treatments for boundary
conditions for this kind of singular potentials, in the frame of
the construction of self adjoint extensions for the Dirac
Hamiltonian \cite{dittrich}.

In this work we use the relativistic partial wave expansion method
with the boundary conditions mentioned above, making a simple
analysis for the scattering states. Moreover, we will also
introduce a double delta potential of the form $V(r)=\pm
a_{1}\delta(r-r_{1})\pm a_{2}\delta(r-r_{2})$.

The idea is to discuss the behavior of the differential and total
cross section when we vary the energy of the incident particles,
analyzing the occurrence of certain resonances in different
channels. We also find a useful relationship between the phase
shifts and the total cross section.

This work is complemented with the use of the Born approximation, that gives us a more general perspective of this problem.

\section{Relativistic scattering}

As it is extremely well known, asymptotically, we expect that the
wave function in the non relativistic case will behave as
\begin{equation}
\Psi(r)\textrm{\tiny $r\tilde{\rightarrow}\infty$} \exp{(iKz)}+f(\theta,\phi)\frac{\exp{(iKr)}}{r}.
\end{equation}
where $K$ is the momentum. The differential cross section is then
\begin{equation}
I(\theta,\phi)=\frac{d\sigma}{d\Omega}=|f(\theta,\phi)|^{2}.
\end{equation}

For scattering at high energies of  1/2-spin particles,
we must consider the Dirac equation
\begin{equation}
(c\hat{\alpha}\cdot\hat{p}+\hat{\beta}mc^{2}+V(r))\psi(r)=H\psi(r),
\end{equation}
where $\hat{\alpha}$ and $\hat{\beta}$ are the usual $4x4$ Dirac matrices, and $\psi$ is the four-component Dirac spinor.

We know that for a free particle, the one dimensional solution of
the above equation is
\begin{eqnarray}
\Psi=\sqrt{\frac{\epsilon+mc^{2}}{2\epsilon}}\left(\begin{array}{c}\chi_{\sigma}\\ \frac{\sigma_{z}p}{\epsilon+mc^{2}}\chi_{\sigma}\end{array}\right)e^{i(p_{z}z-\epsilon t)},
\end{eqnarray}
where $\epsilon=\pm E_{p}$, $\sigma=\pm1$,
$\chi_{1}=\left(\begin{array}{c}1\\0\end{array}\right)$,
$\chi_{-1}=\left(\begin{array}{c}0\\1\end{array}\right)$ and
$\sigma_{z}=\left(\begin{array}{cc}1&0\\0&-1\end{array}\right)$.

When working with a spherical symmetric potential it is better to
transform the Dirac equation into spherical coordinates
\cite{greiner,rose}, where the complete set of commuting
operators is given by $H$, $J^{2}$, $J_{3}$ and $\kappa$, where
$\hat{J}$ is the total angular momentum and  $\hat{\kappa}$ is
defined by
\begin{equation}
\hat{\kappa}=\beta(\frac{2}{\hbar}\hat{S}\cdot\hat{L}+\hbar1).
\end{equation}

The eigenvalues of the operator $\kappa$ are given by
\begin{equation}
\hat{\kappa}\Psi=-\kappa\hbar\Psi=\pm(j+\frac{1}{2})\hbar\Psi,
\end{equation}

If we parametrize the four-component spinor by separating the radial and angular dependence according to the Ansatz
\begin{equation}\label{ansatz}
\Psi=\left(\begin{array}{cc}g(r)&\chi^{\mu}_{\kappa}(\theta,\phi)\\
 if(r)&\chi^{\mu}_{-\kappa}(\theta,\phi)\end{array}\right),
 \end{equation}
where
\begin{equation}
\chi^{\mu}_{\kappa}=\sum_{m}C(l\frac{1}{2}j;\mu-m;m)Y^{\mu-m}_{l}\chi^{m},
\end{equation}
being $C(l\frac{1}{2}j;\mu-m;m)$ are the appropriate Glebsch-Gordan coefficients.
It is easy to see that the  $\hat{\kappa}$ operator satisfies the
following relationship with the angular momentum-spin function
\begin{eqnarray}\label{kappa}
\hat{\kappa}\chi^{\mu}_{\kappa}=-\kappa\chi^{\mu}_{\kappa},\nonumber\\
\hat{\kappa}\chi^{\mu}_{-\kappa}=\kappa\chi^{\mu}_{-\kappa}.
\end{eqnarray}

For the two different coupling, $j=l+1/2$ and $j=l-1/2$, the eigenvalues are given by
\begin{eqnarray}\label{volver}
\kappa=\left\{\begin{array}{lll}
l&j=l-1/2&\kappa>0\\
-l-1&j=l+1/2&\kappa<0
\end{array}.\right.
\end{eqnarray}

The dependence of $\kappa$ on the eigenvalues of the angular momentum will be extremely important for our calculations. Everything will be function of $\kappa $. However, in the future we will replace $\kappa$ with the respective $l_{\kappa}$ remembering the existence of the LS coupling. We notice that:
\begin{eqnarray}
l_{\kappa}=\left\{\begin{array}{ll}
\kappa&\kappa>0\\
-\kappa-1&\kappa<0
\end{array}.\right.
\end{eqnarray}
and
\begin{eqnarray}\label{lbarra}
l_{-\kappa}=\left\{\begin{array}{ll}
\kappa-1&\kappa>0\\
-\kappa&\kappa<0
\end{array}.\right.
\end{eqnarray}

The Dirac equation in spherical coordinates has the form
\begin{equation}\label{ecdirac}
\left(-ic\hat{\alpha}_{r}\left(\frac{\partial}{\partial
r}+\frac{1}{r}-\frac{\hat{\beta}}{r}\hat{\kappa}\right)+\hat{\beta}mc^{2}+V(r)\right)\Psi(r)=E\Psi(r).
\end{equation}

The solution of this equation for a free particle with the Ansatz given in (\ref{ansatz}) is:
\begin{equation}\label{solucion}
G(r)=r(a_{1}j_{l_{\kappa}}(r)+b_{1}n_{l_{\kappa}}(r)),
\end{equation}
\begin{equation}
F(r)=\frac{\kappa}{|\kappa|}\frac{k\hbar
c}{E+mc^{2}}r(a_{1}j_{l_{-\kappa}}(r)+b_{1}n_{l_{-\kappa}}(r)).
\end{equation}
where $G(r)=rg(r)$, $F(r)=rf(r)$ and $j_{l}$, $n_{l}$ are the regular and irregular Bessel functions respectively.

\section{Partial wave expansion method}

Here we will present the relativistic wave expansion method. The reader can see the details in \cite{dapor}. This method shows us how to calculate the differential and total cross section just by knowing the phase shift of the scattered wave.

Let us consider the asymptotic  behavior of a 1/2-spin scattered wave
\begin{equation}
\begin{array}{cc}\Psi_{i}=a_{i}\exp{(iKz)}+b_{i}(\theta,\phi)\frac{\exp{(iKr)}}{r}&i=1,...,4\end{array}.
\end{equation}

To be able to find the effects of the scattering in the spin of the particle, we must first notice that not all the components of the spinor that represents the free particle are independent, which, in turns, allows us to see that not all the $b_{i}(\theta,\phi)$ components are independent.
For example if we have a spin parallel to the direction of incidence (spin up), we have:
\begin{equation}\Psi_{1}\textrm{\tiny $r\tilde{\rightarrow}\infty$}\exp{(iKz)}+f^{+}(\theta,\phi)\frac{\exp{(iKr)}}{r},\end{equation}
\begin{equation}
\Psi_{2}\textrm{\tiny $r\tilde{\rightarrow}\infty$}g^{+}(\theta,\phi)\frac{\exp{(iKr)}}{r}.
\end{equation}

Now, if the spin is anti-parallel to the direction of incidence (spin down) we have
\begin{equation}
\Psi_{1}\textrm{\tiny $r\tilde{\rightarrow}\infty$}g^{-}(\theta,\phi)\frac{\exp{(iKr)}}{r},
\end{equation}
\begin{equation}
\Psi_{2}\textrm{\tiny $r\tilde{\rightarrow}\infty$}\exp{(iKz)}+f^{-}(\theta,\phi)\frac{\exp{(iKr)}}{r}.
\end{equation}

The functions $f^{\pm}(\theta,\phi)$ and $g^{\pm}(\theta,\phi)$ are called \emph{scattering amplitudes} (similar to the non relativistic case). With these amplitudes we can calculate the differential cross section for a non polarized beam, making the change $f^{+}=f^{-}=f$, $g^{+}=g \exp{(i\phi)}$ and $g^{-}=-g \exp{(-i\phi)}$ getting
\begin{equation}\label{sigmadiferencial}
I(\theta,\phi)=\frac{d\sigma}{d\Omega}=|f|^{2}+|g|^{2},
\end{equation}
where the new scattering amplitudes are given by
\begin{equation}
f(\theta)=\sum_{l=0}^{\infty}\hat{A}_{l}P_{l}(\cos{\theta}),
\end{equation}
\begin{equation}
g(\theta)=\sum_{l=1}^{\infty}\hat{B}_{l}P^{1}_{l}(\cos{\theta}),
\end{equation}
and
\begin{equation}\label{al}
\hat{A}_{l}=\frac{1}{2iK}\{(l+1)[\exp{(2i\eta^{+}_{l})}-1]+l[\exp{(2i\eta^{-}_{l})}-1]\},
\end{equation}
\begin{equation}\label{bl}
\hat{B}_{l}=\frac{1}{2iK}[\exp{(2i\eta^{-}_{l})}-\exp{(2i\eta^{+}_{l})}].
\end{equation}

We can see that the differential cross section has a direct dependence on the phase shifts of the wave function. The notation $\eta_{l}^{\pm}$ gives us the dependence of the wave function on the angular momentum and on the two different coupling ($j=l\pm1/2$). It is easy to calculate the total cross section, just by integrating the differential cross section, obtaining
\begin{equation}\label{sigamtotal}
\sigma_{total}=4\pi\sum_{l}\frac{|\hat{A}_{l}|^{2}+l(l+1)|\hat{B}_{l}|^{2}}{2l+1}.
\end{equation}

\section{1-Delta potential}

As we mentioned before, we will apply the wave expansion method to a spherical potential given by
\begin{equation}
V(r)=\pm a\delta(r-r_{0}).
\end{equation}

To be able to find the phase shifts, in order to obtain the differential and total cross section, we must use the boundary conditions described in \cite{loewe}:
\begin{equation}\label{condeborde1}
F^{2}_{+}+G^{2}_{+}=F^{2}_{-}+G^{2}_{-},
\end{equation}
and
\begin{equation}\label{condeborde2}
\frac{F_{+}}{G_{+}}=\frac{(F_{-}/G_{-})+\alpha}{1-\alpha(F_{-}/G_{-})},
\end{equation}
where $F_{+}$ y $G_{+}$ are the solutions of the Dirac equation outside the potential ($r>r_{0}$), and $F_{-}$ y $G_{-}$ are the solutions inside the potential ($r< r_{0}$). $\alpha$ is an dimensionless constant that involves the constant  $a$, $\alpha\equiv \tan{(a/\hbar c)}$.

The first boundary condition tells us that the absolute value of a function with real part $F$ and imaginary part $G$ is constant, which is nothing but the continuity of the probability density. The second boundary condition will be essential to calculate the phase shifts.

First we must find the solution of the Dirac equation in all regions of space. Since we are dealing with a delta type potential, it will only influence the wave function in the support $r=r_{0}$, being, therefore, the solution is the same as in the  free particle case. Separating the space in two regions:

I.- For $r<r_{0}:$
\begin{equation}
G_{1}(r)=ra_{1}j_{l_{\kappa}}(kr),
\end{equation}
\begin{equation}
F_{1}(r)=\frac{\kappa}{|\kappa|}\frac{k\hbar
 c}{E+mc^{2}}a_{1}j_{l_{-\kappa}}(kr).
\end{equation}

The functions $n_{l_{\kappa}}(kr)$ and $n_{l_{-\kappa}}(kr)$ do
not appear in the above equation, because the wave function must remain finite at the origin.

II.- For $r>r_{0}$
\begin{equation}
G_{2}(r)=r(j_{l_{\kappa}}(kr)\cos(\eta^{\pm}_{l})-n_{l_{\kappa}}(kr)sen(\eta^{\pm}_{l})),
\end{equation}
\begin{equation}
F_{2}(r)=\frac{\kappa}{|\kappa|}\frac{k\hbar
c}{E+mc^{2}}(j_{l_{-\kappa}}(kr)\cos(\eta^{\pm}_{l})-n_{l_{-\kappa}}sen(\eta^{\pm}_{l})).
\end{equation}

Here we have written $a_{2}$ and $b_{2}$ as the phase shifts of the scattered wave.
These phase shifts can be found using the boundary conditions. Equation (\ref{condeborde1}) does not give us much valuable information. It only allows us to find the constant $a_{1}$ in the wave functions.  We will concentrate on the second boundary condition (\ref{condeborde2}). Here the constant $a_{1}$ disappear, leaving us $\eta^{\pm}_{l}$ as the only variable.
A simple algebra gives us:
\begin{equation}\label{taneta}
\tan(\eta^{\pm}_{l})=\frac{\alpha(A^{2}j^{2}_{l_{-\kappa}}+j^{2}_{l_{\kappa}})}
{A(n_{l_{-\kappa}}j_{l_{\kappa}}-n_{l_{\kappa}}j_{l_{-\kappa}})+\alpha(n_{l_{\kappa}}j_{l_{\kappa}}+A^{2}n_{l_{-\kappa}}j_{l_{-\kappa}})},
\end{equation}
where
\begin{equation}
A=\frac{\kappa}{|\kappa|}\frac{k\hbar c}{E+mc^{2}}.
\end{equation}

The spherical Bessel functions can be obtained through the
following recurrence relations \cite{gray} :
\begin{equation}
j_{l}(\rho)=\rho^{l}\left(-\frac{1}{\rho}\frac{d}{d\rho}\right)^{l}\left(\frac{\sin{\rho}}{\rho}\right),
\end{equation}
\begin{equation}
n_{l}(\rho)=-\rho^{l}\left(-\frac{1}{\rho}\frac{d}{d\rho}\right)^{l}\left(\frac{\cos{\rho}}{\rho}\right).
\end{equation}

\begin{figure}
\begin{center}
\includegraphics{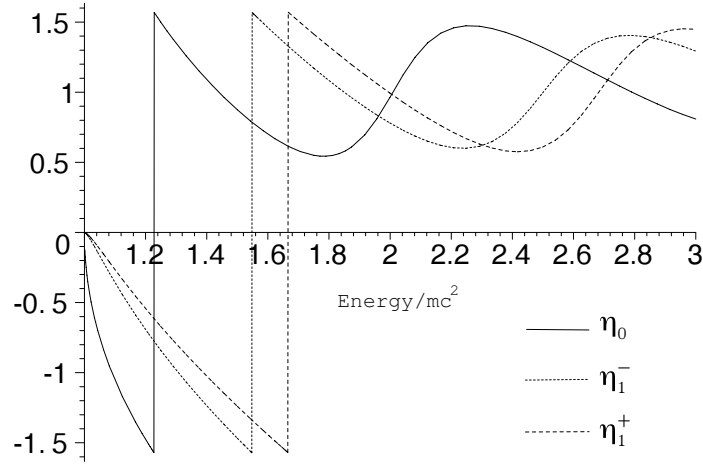}
\caption{Behaviour of the first three
phase shifts for different values of the energy in the one delta
potential case (in units of $mc^{2}$), $\eta_{0}$ (solid curve),
$\eta_{1}^{-}$ (dotted curve), $\eta_{1}^{+}$ (dashed curve). The
super $\pm$ index corresponds to the different angular
momentum-spin couplings, the sub index indicates the angular
momentum channel. Here we use $a=-1\hbar c$ and $r=\hbar/mc$}
\end{center}
\end{figure}

The undulatory behavior for the phase shifts in Figure 1 shows the
typical dependence on the Bessel functions.
Notice the occurrence of discontinuities in each angular
momentum channel, for certain value of energy. This discontinuities, which
 we will identify as resonances, appear when the $\eta^{\pm}_{l}$ get the value $n\pi/2$, with $n$ a natural number. In our work we will only discuss the case where $n=1, -1$. However, we will see that this resonances in the phase shifts, will not affect the cross section.
 
Now, using the results of the phase shifts, we can find an expression for the differential and total cross section using the equations (\ref{sigmadiferencial}) and (\ref{sigamtotal}) for different values of the angular momentum.

Figure 2 shows the total cross section for different values of the energy. Here $l=1$ is the sum of $l=0,1$ and $l=2$ is the sum of $l=0,1,2$ (see equation (\ref{sigamtotal})).

\begin{figure}
\begin{center}
\includegraphics{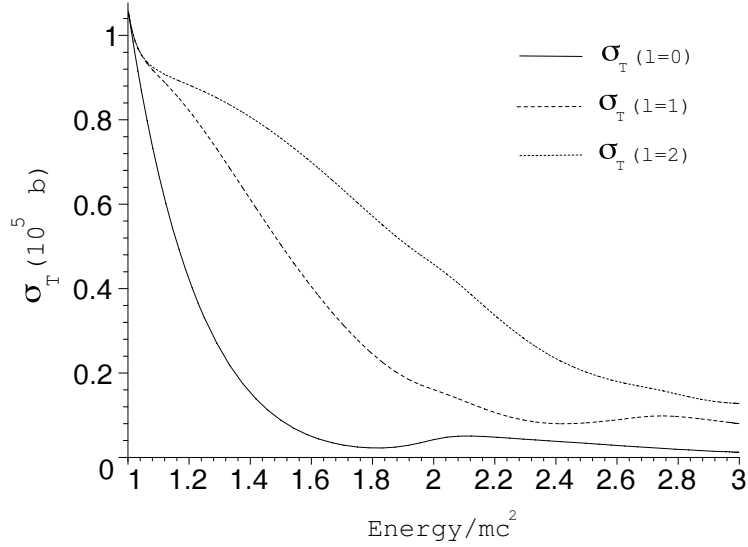}
\caption{Behaviour of the total cross section for different values of the energy in the one delta potential case (in units of $mc^{2}$), $\sigma_{T}(l=0)$ (solid curve), $\sigma_{T}(l=1)$ (dashed curve), $\sigma_{T}(l=2)$ (dotted curve). Here we use $a=-1\hbar c$ and $r=\hbar/mc$ } \end{center} \end{figure}

We first notice that those resonances in the phase shift analysis
do not seem seem to have any consequences in the total cross
section, in spite on their direct influence in equation
(\ref{sigamtotal}). This has to do with the fact that the
resonances in the phase shifts appear at different values of
energies. The cross section vanishes smoothly for high energies.
For low energies (near $mc^{2}$) the dominant channel of the
angular momenta is $l=0$ (as expected). It is also important to
mention that if we increase the delta radius or the coupling
constant $a$ from the potential, the total cross section decreases
and vanishes faster when we increase the energy.

\section{2-Delta potential}

Let us consider a two concentric delta potential:
\begin{equation}V(r)=\pm a_{1}\delta(r-r_{1})\pm
a_{2}\delta(r-r_{2}).\end{equation}

We will carry out the same procedure as before. We separate the space into three regions, finding the corresponding solution of the Dirac equation. Our potential will only affect us when $r=r_{1},r_{2}$ and, hence, the solutions are the same as for a free particle:

1) For $r<r_{1}$
\begin{equation}
G_{1}(r)=ra_{1}j_{l_{\kappa}}(kr),
\end{equation}
\begin{equation}
F_{1}(r)=\frac{\kappa}{|\kappa|}\frac{k\hbar c}{E+mc^{2}}a_{1}j_{l_{-\kappa}}(kr).
\end{equation}

2) For $r_{1}<r<r_{2}$
\begin{equation}
G_{2}(r)=r(a_{2}j_{l_{\kappa}}(kr)+b_{2}n_{l_{\kappa}}(kr)),
\end{equation}
\begin{equation}F_{2}(r)=\frac{\kappa}{|\kappa|}\frac{k\hbar c}{E+mc^{2}}(a_{2}j_{l_{-\kappa}}(kr)+b_{2}n_{l_{-\kappa}}(kr)).
\end{equation}

3) For $r>r_{2}$
\begin{equation}
G_{3}(r)=r(j_{l_{\kappa}}(kr)\cos{\eta^{\pm}_{l}}-n_{l_{\kappa}}(kr)\sin{\eta^{\pm}_{l}}),
\end{equation}
\begin{equation}F_{3}(r)=\frac{\kappa}{|\kappa|}\frac{k\hbar
c}{E+mc^{2}}(j_{l_{\kappa}}(kr)\cos{\eta^{\pm}_{l}}-n_{l_{-\kappa}}(kr)\sin{\eta^{\pm}_{l}}).
\end{equation}

Using the boundary conditions for each delta and having in mind that $\alpha_{1}=\tan{(a_{1}/\hbar c)}$ and $\alpha_{2}=\tan{(a_{2}/\hbar c)}$, we can find an expression for the phase shifts:
\begin{equation}\label{pikoro}
\tan{\eta^{\pm}_{l}}=\frac{\alpha_{2}\left(A^{2}j^{2}_{l_{-\kappa}}+j^{2}_{l_{\kappa}}+A^{2}n_{l_{-\kappa}}j_{l_{-\kappa}}+n_{l_{\kappa}}
    j_{l_{\kappa}}\right)+\tilde{A}(kr_{1})
    \left(A(n_{l_{\kappa}}j_{l_{-\kappa}}-n_{l_{-\kappa}}j_{l_{\kappa}})\right)}{A(n_{l_{-\kappa}}j_{l_{\kappa}}-n_{l_{\kappa}}j_{l_{-\kappa}})+\alpha_{2}\left(A^{2}n_{l_{-\kappa}}
j_{l_{-\kappa}}+n_{l_{\kappa}}j_{l_{\kappa}}+\tilde{A}(kr_{1})(A^{2}n^{2}_{l_{-\kappa}}+n^{2}_{l_{\kappa}})\right)},
\end{equation}
where
\begin{displaymath}A=\frac{\kappa}{|\kappa|}\frac{k\hbar
c}{E+mc^{2}},\end{displaymath}
and
\begin{equation}\label{goku}
\tilde{A}(kr_{1})=\frac{\alpha_{1}\left(A^{2}j^{2}_{l_{-\kappa}}+j^{2}_{l_{\kappa}}\right)}{A(n_{l_{\kappa}}j_{l_{-\kappa}}
-n_{l_{-\kappa}}j_{l_{\kappa}})-\alpha_{1}(n_{l_{\kappa}}j_{l_{\kappa}}+A^{2}n_{l_{-\kappa}}j_{l_{-\kappa}})}.
\end{equation}

In equation (\ref{pikoro}) all Bessel functions depend only on
$kr_{2}$. The dependence on $kr_{1}$ is concentrated on the
function $\tilde{A}(kr_{1})$ as shown on (\ref{goku}). Now, if we
choose $a_{1}=0$ or $r_{1}=r_{2}$ the problem, and hence the
solution, reduces to a one delta potential.

Knowing the phase shifts for a two-delta potential, we can obtain an expression for the total cross section. Figure 3 shows the cross section as a function of the energy.

\begin{figure} 
\begin{center}
\includegraphics{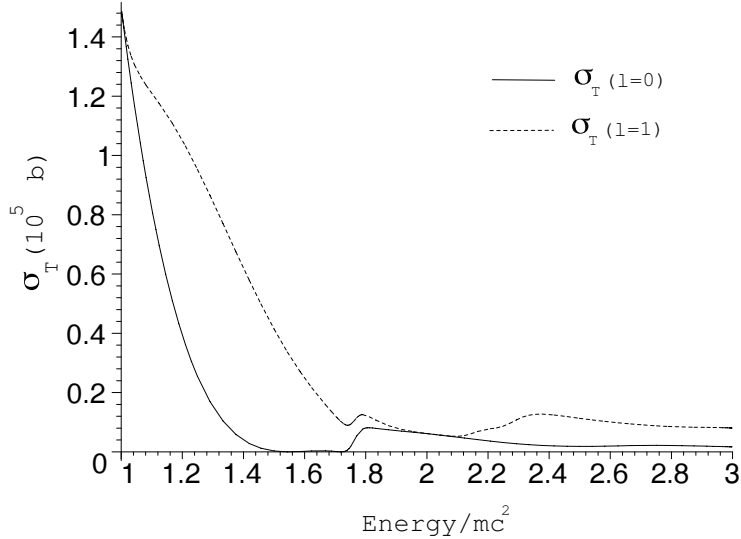} 
\caption{Behaviour of the total cross section for different values of the energy in the two delta potential case (in units of $mc^{2}$), $\sigma_{T}(l=0)$ (solid curve), $\sigma_{T}(l=1)$ (dashed curve). Here we use $a_{1}=1\hbar c$, $a_{2}=1\hbar c$, $r_{1}=2\hbar/mc$ and $r_{2}=3\hbar/mc$ } 
\end{center} 
\end{figure}

Again we can see that the resonances of the phase shifts do not have any notorious influence on the total cross section, in the same way as happened for a one delta potential. It decays smoothly to zero for high energies. For small energies, near $mc^{2}$, the dominant angular momentum is the $l=0$ channel. Once again, we must remember that $l=1$ stands for the sum of $l=0,1$, and $l=2$ stands for the sum of $l=0,1,2$.

\section{Born Approximation}

Let us remember our potential
\begin{equation}
V(r)=\pm a_{1}\delta(r-r_{1})\pm
a_{2}\delta(r-r_{1}).
\end{equation}

Following the same steps shown in Itzykson-Zuber \cite{born}
(equation 2-126), we get for the scattering amplitude:
\begin{equation}
S_{fi}=\frac{im}{V(E_{i}E_{f})^{\frac{1}{2}}}(2\pi)\delta(E_{f}-E_{i})
\int{d^{3}r\exp({-i\vec{q}\cdot\vec{r})}V(r)\bar{u}^{\alpha}(p_{f})\gamma^{0}u^{\beta}(p_{i})},
\end{equation}
where $\vec{q}=\vec{p}_{f}-\vec{p}_{i}$. Then
$\vec{r}\cdot\vec{r}=|\vec{q}||\vec{r}|\cos{\theta}=|\vec{q}||\vec{r}|\mu$,
with $\mu=\cos{\theta}$.

Integrating in spherical coordinates: $\int^{1}_{-1}d\mu\int^{2\pi}_{0}d\phi\int^{\infty}_{0}r^{2}dr$,
defining
$\Gamma\equiv\bar{u}^{\alpha}(p_{f})\gamma^{0}u^{\beta}(p_{i})$,
we obtain
\begin{equation}
S_{fi}=\frac{im}{V(E_{i}E_{f})^{\frac{1}{2}}}\frac{4\pi^{2}}{q}
\delta(E_{f}-E_{i})[\pm a_{1}r_{1}\sin{qr_{1}}\pm a_{2}r_{2}\sin{qr_{2}}]\Gamma.
\end{equation}

The transition probability from $i$ to $f$ per time and particle
is:
\begin{eqnarray}
\frac{dP_{fi}}{dt}&=\int\left|\frac{im}{V(E_{i}E_{f})^{\frac{1}{2}}}\frac{2\pi}{|\vec{q}|}(\pm
a_{1}r_{1}\sin{qr_{1}}\pm
a_{2}r_{2}\sin{qr_{2}})\Gamma\right|^{2}\nonumber\\
&\times2\pi\delta(E_{f}-E_{i})V\frac{d^{3}p_{f}}{(2\pi)^{3}}.
\end{eqnarray}

A  straightforward calculation gives us
\begin{equation}\label{popopota}
\frac{dP_{fi}}{dt}=\int\frac{m^{2}}{V(E_{i}E_{f})}\frac{(\pm a_{1}r_{1}\sin{qr_{1}}\pm
a_{2}r_{2}\sin{qr_{2}})}{|\vec{q}|}|\Gamma|^{2}
\delta(E_{f}-E_{i})d^{3}p_{f}.
\end{equation}

To be able to find the differential cross section, we must divide
the last equation by the incident flux
$\frac{1}{V}\frac{|p_{i}|}{E_{i}}$
\begin{equation}
d\sigma=\int\frac{m^{2}}{|p_{i}|E_{f}|q|^{2}}(\pm a_{1}r_{1}\sin{qr_{1}}\pm
a_{2}r_{2}\sin{qr_{2}})^{2}|\Gamma|^{2}
\delta(E_{f}-E_{i})p^{2}_{f}dp_{f}d\Omega.
\end{equation}

Using the fact that $|p_{i}|=|p_{f}|=p_{f}$ and
$p_{f}dp_{f}=E_{f}dE_{f}$ we obtain
\begin{equation}\frac{d\sigma}{d\Omega}=\frac{m^{2}}{|q|^{2}}(\pm a_{1}r_{1}\sin{qr_{1}}\pm
a_{2}r_{2}\sin{qr_{2}})^{2}|\Gamma|^{2}.\end{equation}

Since we are interested in the cross section for a non polarized
beam, we must sum over all $\alpha$ and $\beta$ that appear on the function $\Gamma$, this means
\begin{equation}
\frac{d\sigma}{d\Omega}=\frac{m^{2}}{|q|^{2}}(\pm a_{1}r_{1}\sin{qr_{1}}\pm
a_{2}r_{2}\sin{qr_{2}})^{2}\sum_{\alpha}\frac{1}{2}\sum_{\beta}|\bar{u}^{\alpha}(p_{f})\gamma^{0}u^{\beta}(p_{i})|^{2}.
\end{equation}

We will concentrate directly on the sum over $\alpha$ and $\beta$.
Rewriting  the sum as traces, we get
\begin{equation}
\sum_{\alpha}\frac{1}{2}\sum_{\beta}|\bar{u}^{\alpha}(p_{f})\gamma^{0}u^{\beta}(p_{i})|^{2}=\frac{1}{2}
tr\left(\gamma^{0}\frac{(\gamma^{0}p_{i})+m}{2m}\gamma^{0}\frac{(\gamma^{0}p_{f})+m}{2m}\right),
\end{equation}
and using some identities for the traces of the  $\gamma$
matrices, we have
\begin{equation}tr(\gamma^{0}(\gamma^{0}p_{i})\gamma^{0}(\gamma^{0}p_{f})=4(E_{i}E_{f}-p_{i}\cdot p_{f}+E_{i}E_{f}),\end{equation}
\begin{equation}
tr(\gamma^{0}\gamma^{0})=4.
\end{equation}

We will also need some kinematics relations
\begin{equation}
E_{i}=E_{f}+E,
\end{equation}
\begin{equation}
p_{i}\cdot p_{f}=E^{2}-p^{2}\cos{\theta}=m^{2}+2E^{2}\beta^{2}\sin^{2}{\frac{\theta}{2}},
\end{equation}
where $\beta\equiv v/c=|p|/E$ is the incoming velocity. In this
way we obtain
\begin{equation}\frac{1}{2}
tr\left(\gamma^{0}\frac{(\gamma^{0}p_{i})+m}{2m}\gamma^{0}\frac{(\gamma^{0}p_{f})+m}{2m}\right)=\frac{E^{2}}{m^{2}}(1-\beta^{2}\sin^{2}{\frac{\theta}{2}}).\end{equation}

The differential cross section with
$|\vec{q}|^{2}=4|\vec{p}|^{2}\sin^{2}{\frac{\theta}{2}}=4\beta
E|\vec{p}|\sin^{2}{\frac{\theta}{2}}$ is:
\begin{eqnarray}\label{borndiferencial}
\frac{d\sigma}{d\Omega}&=\frac{E(1-\beta^{2}\sin^{2}{\frac{\theta}{2}})}{4\beta|\vec{p}|\sin^{2}{\frac{\theta}{2}}}[\pm a_{1}r_{1}\sin{(2\sqrt{\beta E|\vec{p}|}\sin{\frac{\theta}{2}r_{1}})}\nonumber\\
&\pm a_{2}r_{2}\sin{(2\sqrt{\beta
E|\vec{p}|}\sin{\frac{\theta}{2}r_{2}})}]^{2}.
\end{eqnarray}

We can easily see that the differential cross section is the same for attractive or repulsive potentials. One important thing to notice, when working with the partial wave method, we focus on different channels of the angular momentum. But, however, we take the sum of all the angular momenta in the Born approximation.

The total cross section is given by
\begin{equation}\sigma_{total}=\int
d\theta\sin{\theta}d\phi\frac{d\sigma}{d\Omega},\end{equation}
where $d\sigma/d\Omega$ is defined in
(\ref{borndiferencial}), We obtain the following equation:
\begin{eqnarray}\label{integrales144}\sigma_{total}=\frac{2\pi E}{4\beta|\vec{p}|}\int^{\pi}_{0}
d\theta\sin{\theta}[a_{1}^{2}r_{1}^{2}\sin^{2}{(\alpha_{1}\sin{\frac{\theta}{2}})}
+a_{2}^{2}r_{2}^{2}\sin^{2}{(\alpha_{2}\sin{\frac{\theta}{2}})}\nonumber\\
+2a_{1}a_{2}r_{1}r_{2}\sin{(\alpha_{1}\sin{\frac{\theta}{2}})}\sin{(\alpha_{2}\sin{\frac{\theta}{2}})}]
\left(\frac{1-\beta^{2}\sin^{2}{\frac{\theta}{2}}}{\sin^{2}{\frac{\theta}{2}}}\right).\end{eqnarray}
where $\alpha_{1}\equiv 2\sqrt{\beta E|\vec{p}}r_{1}$ y
$\alpha_{2}\equiv 2\sqrt{\beta E|\vec{p}}r_{2}$. We must notice
that the factor $\pm$ is not taken into account. As before we realized that
 there will be no influence on the results if both delta potentials are repulsive or attractive. The case where the sign of $a_{1}$ is different from the sign of $a_{2}$ will also not give new information.
Integrating directly (\ref{integrales144}) we obtain:
\begin{eqnarray}\label{borntotal}\sigma_{total}&&=\frac{\pi
E}{2\beta|\vec{p}|}\{-2Ci(2\alpha_{1})(a_{1}r_{1})^{2}-2Ci(2\alpha_{2})(a_{2}r_{2})^{2}
+2\ln{(2\alpha_{1})}(a_{1}r_{1})^{2}\nonumber\\
&&+2\ln{(2\alpha_{2})}(a_{2}r_{2})^{2}-\beta^{2}((a_{1}r_{1})^{2}+(a_{2}r_{2})^{2})\nonumber\\
&&+\beta^{2}\left(\frac{\sin{(2\alpha_{1})}}{\alpha_{1}})(a_{1}r_{1})^{2}+\frac{\sin{(2\alpha2})}
{\alpha_{2}}(a_{2}r_{2})^{2}\right)\nonumber\\
&&-\beta^{2}\left(\frac{\sin^{2}{(\alpha_{1})}}{\alpha_{1}^{2}}(a_{1}r_{1})^{2}+\frac{\sin^{2}{(\alpha2})}
{\alpha_{2}^{2}}(a_{2}r_{2})^{2}\right)\nonumber\\
&&+4(a_{1}r_{1})(a_{2}r_{2})(Ci(\alpha_{1}-\alpha_{2})-Ci(\alpha_{1}+\alpha_{2})\nonumber\\
&&-\beta^{2}[\frac{\sin{(\alpha_{1}-\alpha_{2})}}{\alpha_{1}-\alpha_{2})}+\frac{\cos{(\alpha_{1}-\alpha_{2})}}
{(\alpha_{1}-\alpha_{2})^{2}}-\frac{1}{(\alpha_{1}-\alpha_{2})^{2}}\nonumber\\
&&-\frac{\sin{(\alpha_{1}+\alpha_{2})}}{(\alpha_{1}+\alpha_{2})}-\frac{\cos{(\alpha_{1}+\alpha_{2})}}
{(\alpha_{1}+\alpha_{2})^{2}}+\frac{1}{(\alpha_{1}+\alpha_{2})^{2}}])\nonumber\\
&&-C(2(a_{1}r_{1})^{2}+2(a_{2}r_{2})^{2}),
\end{eqnarray}
where $Ci$ is the well known cosine integral, given by the following equation:
\begin{equation}Ci(x)=C+\ln{(x)}+\sum^{\infty}_{k=1}(-1)^{k}\frac{x^{2k}}{2k(2k)!},\end{equation}
and C is the Euler constant.

Figure 4 shows the dependence of the total cross section while we
vary the energy of the particles. We notice that the total cross
section has the same behavior found in the partial wave method. On
the other hand if we take a $\Delta r=r_{2}-r_{1}=cte$, we can see
that the behavior is again similar to  the one we found in the
partial wave method.

\begin{figure}
 \begin{center}
  \includegraphics{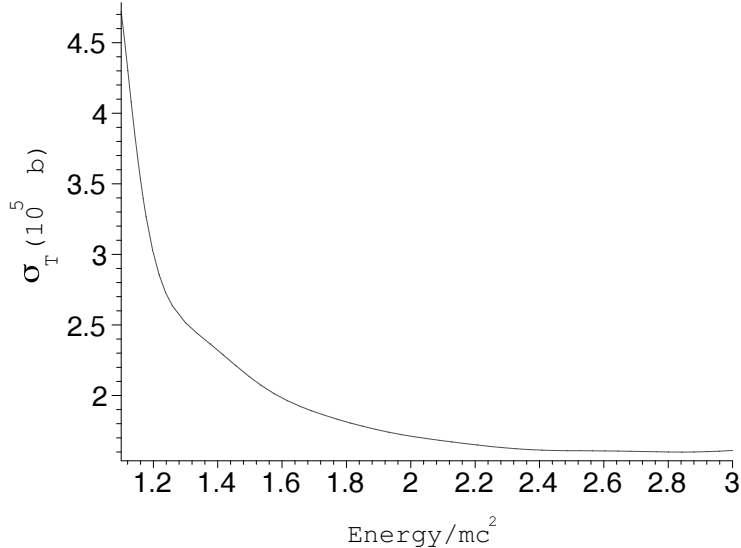} 
  \caption{Behaviour of the total cross section for different values of the energy in the two delta potential case (in units of $mc^{2}$) with the Born Approximation. Here we use $a_{1}=-1\hbar c$, $a_{2}=-1\hbar c$, $r_{1}=2\hbar/mc$ and $r_{2}=3\hbar/mc$ } 
  \end{center}
   \end{figure}

We can say that the behavior of the cross section given by the sum
of all the angular momenta is similar to the behavior of the cross
section for a specific angular momentum channel.
In spite of several efforts, we were unable to find geometric
conditions related to the separation of the two delta potentials,
and the possible occurrence of resonances as peaks in the total
cross section.

\section{Conclusions}

The general idea of this work was to use the boundary conditions
obtained in \cite{loewe}, when working with delta type singular
potentials, in particular for a spherical potential like $V(r)=\pm a\delta(r-r_{0})$ in the scattering region. We studied the behavior of the phase shifts, the differential and total cross section.

We found a direct relation between the phase shifts of the scattered wave
and the cross section. With this we were able to calculate values of the cross section for different angular momentum channels, giving us valuable information, in particular, the dependence of the cross section on the $l=0$ channel  in the low energy limit.

It is important to notice that the behavior of the cross sections for each channel of the angular momenta is very similar for repulsive and attractive potentials, in the one and two delta cases.
We confirmed our results through the calculation of the cross section within the Born approximation. This method does not require the knowledge of the boundary conditions, but it does not give us information about the angular momenta of the particles.

It could be interesting to expand this work to an anion-potential,
i.e. a series of concentric delta-potentials, trying to find
recurrence relations for the occurrence of resonances in the
different angular momentum channels.

\begin{acknowledgments}
We acknowledge financial support from Fondecyt 1010976
\end{acknowledgments}

\end{document}